\documentclass{an}
\usepackage{graphicx}
\usepackage{times}
\usepackage{fancyhdr}
\def\part{\partial}

\def\bfA{{\bf A}}
\def\bfB{{\bf B}}

\def\beqn{\begin{eqnarray}}
\def\eeqn{\end{eqnarray}}

\def\.{\mathaccent 95}
\def\beq{\begin{equation}}
\def\ee{\end{equation}}

\def\ep{\epsilon}

\def\frac#1#2{{\textstyle{{#1}\over {#2}}}}

\def\lsim{\mathrel{\rlap{\lower4pt\hbox{\hskip1pt$\sim$}}
    \raise1pt\hbox{$<$}}}
\def\gsim{\mathrel{\rlap{\lower4pt\hbox{\hskip1pt$\sim$}}
    \raise1pt\hbox{$>$}}}
\def\sqr#1#2{{\vcenter{\vbox{\hrule height.#2pt
         \hbox{\vrule width.#2pt height#1pt \kern#1pt
         \vrule width.#2pt}
         \hrule height.#2pt}}}}

\newbox\grsign \setbox\grsign=\hbox{$>$} \newdimen\grdimen \grdimen=\ht\grsign
\newbox\simlessbox \newbox\simgreatbox
\setbox\simgreatbox=\hbox{\raise.5ex\hbox{$>$}\llap
     {\lower.5ex\hbox{$\sim$}}}\ht1=\grdimen\dp1=0pt
\setbox\simlessbox=\hbox{\raise.5ex\hbox{$<$}\llap
     {\lower.5ex\hbox{$\sim$}}}\ht2=\grdimen\dp2=0pt

\def\doublespace {\smallskipamount=6pt plus2pt minus2pt
                  \medskipamount=12pt plus4pt minus4pt
                  \bigskipamount=24pt plus8pt minus8pt
                  \normalbaselineskip=24pt plus0pt minus0pt
                  \normallineskip=2pt
                  \normallineskiplimit=0pt
                  \jot=6pt
                  {\def\smallskip {\vskip\smallskipamount}}
                  {\def\medskip   {\vskip\medskipamount}}
                  {\def\bigskip   {\vskip\bigskipamount}}
                  {\setbox\strutbox=\hbox{\vrule 
                    height17.0pt depth7.0pt width 0pt}}
                  \parskip 12.0pt
                  \normalbaselines}

\font\gkvec=cmmib10                         
\def\bomega{\hbox{{\gkvec\char33}}}                  

\def\lb{\langle}
\def\rb{\rangle}

\def\bw{\bar{\omega}}
\def\bv{\bar V}
\def\bB{\overline B}
\def\ts{\times}
\def\lb{\langle}
\def\rb{\rangle}
\def\curl{\nabla {\ts}}

\def\bfv{{\bf v}}

\def\bfj{{\bf j}}

\def\bfw{{\bomega}}

\def\bfb{{\bf b}}

\def\bfB{{\bf B}}

\def\bbB{\overline {\bf B}}
\def\nb{\nabla}
\def\curl{\nb\ts}

\def\b0{b^{(0)}}
\def\v0{v^{(0)}}
\def\w0{\omega^{(0)}}
\def\bb0{\bfb^{(0)}}
\def\bv0{\bfv^{(0)}}
\def\bw0{\bfw^{(0)}}
\def\bj0{\bfj^{(0)}}

\begin{document}

\title{On the Meaning and Inapplicability of the 
Zeldovich Relations of Magnetohydrodynamics}

\author{Eric G. Blackman\inst{1} \and  George B. Field\inst{2}}
\institute{Department of Physics and Astronomy, University of Rochester,
Rochester NY, 14627, USA 
\and
Center for Astrophysics, 60 Garden St., Cambridge MA, 02139, USA}

\date{Received; accepted; published online}

\abstract{Considering a plasma with an initially weak
 large scale field 
subject to nonhelical turbulent stirring, 
Zeldovich (1957), for two-dimensions, followed by others
for three dimensions,  
have presented formulae of the form 
$\lb{b}^2\rb=f(R_M){\bB}^2$.  Such ``Zeldovich relations'' 
have sometimes been interpreted to provide steady-state relations 
between the energy associated with 
the fluctuating magnetic field  
and that associated with a large scale or mean field multiplied by  a
function $f$ that 
depends on  spatial dimension and a magnetic Reynolds number $R_M$. 
Here we dissect the origin of these relations  and 
pinpoint pitfalls  that show
why  they are inapplicable to realistic, dynamical  
MHD turbulence and that they disagree with many numerical simulations.  
For 2-D, we show that when the total magnetic field is 
determined by a vector potential, the standard Zeldovich relation 
applies only  transiently, characterizing a maximum possible   
value that the field energy can reach before necessarily decaying.
in relation to a seed value $\bB$.
In 3-D, we show that the standard Zeldovich relations 
are derived by balancing  subdominant  terms.  
In contrast, balancing the dominant terms shows that 
the fluctuating field can grow 
to a value independent of $R_M$ and the initially imposed $\bB$, 
as seen in numerical simulations. 
We also emphasize that these Zeldovich relations of nonhelical turbulence  
imply nothing about the amount  mean field growth in a helical dynamo.
In short, by re-analyzing the origin of the Zeldovich relations, we highlight 
that they are inapplicable to realistic steady-states of large $R_M$ 
MHD turbulence.
\keywords{galaxies: magnetic fields---
ISM: magnetic fields---stars: magnetic fields}}

\correspondence{blackman@pas.rochester.edu}

\maketitle

\section{Introduction}
Stars, galactic interstellar media, the hot plasma   
of galaxy clusters, and accretion disks 
all contain magnetohydrodynamic (MHD) turbulence.
Over the past 50 years, the ubiquity of MHD turbulence
and magnetic fields in astrophysics has stimulated an effort to understand 
the resulting magnetic
and kinetic energy spectra (e.g. Biskamp 1997; 2003; Brandenburg \& Subramanian 2005). 

Simplified semi-analytic models combined with 
idealized numerical experiments with restricted boundary conditions
have been used in combination, 
to help bridge  
the gap between  ignorance and  realistic astrophysical 
MHD turbulent settings. 
One such class of studies focuses on the 3-D
amplification of a weak seed magnetic field by a nonhelical 
turbulent flow 
in a closed volume or periodic box (e.g. Kazanstev 1968, Kulsrud \& Anderson 1992; Maron \& Blackman 2002; 
Haugen, Brandenburg, Dobler 2003; Haugen, Brandenburg, Dobler 2004;
Maron, Cowley, McWilliams 2004; Schekochihin et al. 2004).
Starting with an 
initial seed field, the system is forced with an 
incompressible  turbulent 
flow. The key questions are:  
how much magnetic energy grows and  what determines its saturated
value and spectrum?  Two initial value problems can be distinguished:
(1) An  initially weak (magnetic energy $\ll$ turbulent forcing energy)
seed magnetic field with a finite power  at 
 wavenumber $k=0$, subject to forced, non-helical isotropic
turbulence at forcing wavenumber $k_{\rm f}\sim 5k_1$ (e.g. Haugen \& Brandenburg 2004)
(2) An  initially weak seed random magnetic field with no 
$k=0$ field, subject to forced, non-helical isotropic
turbulence at $k_{\rm f}\sim 5$ (e.g. 
Maron et al. 2004; Schekochihin et al. 2004).

The two initial value problems  above 
address the 3-D small scale dynamo, defined as  field amplification
at or below the forcing scale.
Simulations show that for nonhelical randomly forced flows, 
no significant field amplification above the scale of the forcing 
occurs 
(e.g. Maron \& Blackman 2002; 
Haugen et al. 2003, 2004;  Maron et al. 2004).
When a $k=0$  field is imposed and a numerical simulation is performed
with periodic box, this mean field cannot change due to the boundary
conditions. 
If the turbulence is instead forced helically at $k=k_{\rm f}$, 
the $k=0$ field still cannot grow, but the field up to $k= 1$ can 
and does grow (e.g. Pouquet, Leorat \& Frisch 1976; Brandenburg 2001; Maron \& Blackman 2002; Blackman \& Field 2002).
(The explicit difference between the helical and non-helical forcing
can be seen in the set of simulations by Maron \& Blackman (2002).)
Any significant field amplification on scales above the forcing scale,
involves the large scale dynamo mechanism, and requires helical forcing. 
The nonlinear helical dynamo is the dynamical 
generalization of the original large scale 
dynamo (e.g. Moffatt 1978; Parker 1979) proposed 
to model the large scale fields of rotating stratified systems like 
the sun or Galaxy.

The present focus is on the nonhelical dynamo saturation.
Two key results emerging from 3-D  numerical simulations of the nonhelical
dynamo are that (1) no significant field amplification occurs
at $k<k_{\rm f}$ and (2) 
the total turbulent magnetic energy integrated for $k\ge k_{\rm f}$
grows to within a factor 
of a few of equipartition with the incompressible turbulent kinetic energy
(when the magnetic Prandtl number $Pr_M\equiv \nu/\lambda$ (where
$\nu$ is the viscosity and $\lambda$ is the magnetic diffusivity)
 is not too small;  Haugen et al. 2004; Boldyrev \& Cattaneo 2004. This is independent of whether there is an initial $k=0$, or other large scale $(0<k<k_{\rm f})$ field present, 
as long as that initial  large scale 
field is weak compared to the turbulent 
forcing (e.g. Haugen and Brandenburg 2004). In fact, the saturated
end state of  the nonhelical MHD turbulent dynamo spectrum is essentially independent of the
shape of the seed spectrum (Maron et al.  2004).

The approximate equipartition and its independence of an initially weak 
large scale field  seem physically reasonable,
but  contradict a classic relation in MHD referred to as the ``Zeldovich
relation.'' This refers to the Zeldovich (1957, hereafter Z57) 
result derived for 2-D, and subsequently generalized 
for 3-D using different approaches (e.g. Steenbeck \& Krause 1969; Low 1972; Parker 1973; 
Krause \& R\"adler 1980; Zeldovich, Ruzmaikin, Sokoloff 1983 (hereafter  Z83)).
 The Zeldovich relation is given by  
\beq
\lb{\bfb}^2\rb=f(R_M){\bbB}^2,
\label{a1}
\ee
where the magnetic field is  
$\bfB=\bfb +\bbB$, with $\bfb$  the fluctuating field and
$\bbB$ the mean  or large scale field. 
Eq. (\ref{a1}) is sometimes used to purport 
a steady state relation for the fluctuating
($\sim$ total) 
magnetic energy  and the magnetic energy in $\bbB$ 
when the latter provides an initial seed. 
(We will see later that the  distinction between $\bbB$ as a
large scale field derivable from a vector potential vs. 
$\bbB$ as a mean $k=0$ field has a particular implication for 2-D.)

The function $f(R_M)$ of the magnetic Reynolds number
$R_M$ is $\propto R_M^{-1}$ for
two dimensions (Z57; Moffatt 1978).
In three dimensions, different values have been proposed analytically:
As discussed in Krause and R\"adler (1980, hereafter KR), for the large $R_M$ limit, 
Steenbeck \& Krause (1969), Low (1972),  Parker (1973)
all find $f(R_M) \sim R_M$, while $f(R_M)\sim R_M^2$ for the low $R_M$ limit.
In contrast, in Z83, a more spectrally sensitive result is proposed.
There $f(R_M)\propto ln (R_M)$ for three dimensions and Kolmogorov turbulence, 
and $\propto R_M^{5-3p\over 3-p}$ in 3-D when the kinetic spectral index $p \ne 5/3$.  

Two issues immediately arise: First, why do the 3-D analytic calculations
differ so strongly in the form of $f(R_M)$? 
Second, having just discussed that modern simulations show
that MHD turbulence saturates to a state in which the 
fluctuating field energy has a value independent
of any large scale field, how can a relation like 
(\ref{a1}) apply regardless of the specific form of $f(R_M)$?
Where do relations of the form (\ref{a1}) come from and how
can we understand why they do not agree with simulated MHD steady states?
We address these questions in the present paper.

Note that before the non-linear regime could be studied with numerical
simulations, 
most dynamo theorists studied the kinematic
regime. This refers to the regime in which the magnetic field
amplified by a turbulent flow does not significantly back-react
on this flow, so that the velocity is not subject to strong Lorentz forces.
This regime lends itself to analytic studies, but 
the limitations of the implications must be properly understood.
The Zeldovich relations were derived from this perspective,
but the specific reasons why different derivations appear to disagree 
and why none are applicable to the steady-state of fully developed MHD
turbulence have not been succinctly elucidated. Doing so 
is our present goal.

In section 2, we discuss the Zeldovich relation in 2-D,
and explain that it cannot represent a steady state if the total
field is derived from a vector potential.
We show instead that it represents an absolute maximum that the fluctuating
field can attain before necessarily decaying.
In section 3, we show that in 3-D, the Zeldovich  relations have
been  derived using  arguments ignore key nonlinear growth terms and thus
balance subdominant terms. 
We also show that, although the differences
in the 3-D forms of (\ref{a1}) mentioned above 
can be attributed specifically do different
approaches and inclusion of different subdominant terms,
all of the analytical approaches drop the same dominant term.
As a result, the generalized Zeldovich relation for 3-D MHD turbulence is 
inapplicable to  steady-state large $R_M$ MHD turbulence. 
We conclude in section 4.

We emphasize throughout, that the Zeldovich relations were 
always derived in the context of nonhelical turbulence.
They were therefore never meant to apply to a system in which $\bbB$ 
grows, and thus do not present any constraints on helical
dynamo theory. This can  lead to confusion because the 
Zeldovich relation sometimes appears in discussions whose main 
focus is on mean field helical dynamo 
amplification (e.g. Vainshtein \& Cattaneo 1992). 
Our detailed discussion is therefore focused specifically on understanding
the  inapplicability of the Zeldovich relation to  steady-state
 nonhelical MHD turbulence.

\section{Understanding the  2-D Zeldovich Relation}

We first address the 2-D Zeldovich relation derived
by Z57.  Since no dynamo action
can be sustained in 2-D, and the magnetic energy must ultimately
decay, one immediately wonders what a relation between
$\bbB^2$ and $\lb b^2\rb$ could mean?
In fact Z57 is primarily a 2-D anti-dynamo theorem paper: 
we will see that the relation between 
$\bbB$ and $\lb b^2\rb$ represents
the maximum that $\lb b^2 \rb$ 
could possibly attain before it necessarily decays.

We start with a basic argument showing that in 2-D,
when the total magnetic field (mean + fluctuating)
is determined by a vector potential,
(i.e. $\bfB=\curl \bfA$ and no $k=0$ component for a periodic box) 
the total magnetic energy must decay when surface integrals vanish.  
We write the total the 
incompressible induction equation for the total magnetic field   
as (e.g. Moffatt 1978)
\beq
\partial_t{\bf B}= \curl (\bfv\ts\bfB)+\lambda \nabla^2\bfB=
{\bf B}\cdot\nabla{\bfv}-{\bfv}\cdot\nabla{\bf B}+\lambda\nabla^2{\bf B},
\label{z5701}
\ee
where $\bfv$ is the velocity flow, assumed to be imposed by external forcing
with no mean component, and $\lambda$ is the magnetic diffusivity.

Now following Z57 and 
restricting to 2-D $(x,y)$ incompressible flow, we can then separate
out the equation for the $z$-component  
\beq
\partial_t B_z=-{\bfv} \cdot \nabla B_z+\lambda \nabla^2 B_z.
\label{z5702}
\ee
Multiplying by $B_z$ and spatially averaging for incompressible
flows, we have
\beq
\begin{array}{r}
\partial_t\lb B_z^2\rb=-\partial_i\lb v_i B_z^2\rb
+\lambda\partial_j\lb\partial_j B_z^2\rb-2\lambda\lb(\partial_k B_z)^2\rb\\
= -2\lambda\lb(\partial_k B_z)^2\rb, 
\end{array}
\label{z5703}
\ee
where the latter equality follows when we assume
surface integrals vanish. Eq. (\ref{z5703})
shows that the $b_z$ contribution to the 
magnetic energy will decay.

Because of the first term on the right of (\ref{z5701}),
the $x,y$ components of the magnetic field need  not 
immediately decay, although they will eventually decay. To
see this, note that the 
vector potential (defined such that $\bfB\equiv \curl\bfA$) satisfies
\beq
\partial_t{\bf A}= {\bfv}\ts (\curl {\bf A})
+\lambda\nabla^2 {\bf A} -\lambda{\nabla({\nabla\cdot {\bf A}})} -
\nabla\phi.
\label{z5704}
\ee
As above,
 we can  separate out the $z$-component, which determines $B_x$ and $B_y$, 
to obtain
\beq
\partial_t{A_z}= -{\bfv}\cdot \nabla {A_z}
+\lambda\nabla^2 {A_z},
\label{z5705}
\ee
since the $z-$derivatives vanish. Eq. (\ref{z5705})
has the same form as (\ref{z5702}).
Multiplying by $A_z$ and averaging, we again have for incompressible flow
\beq
\begin{array}{r}
\partial_t\lb{A_z}^2\rb=-\partial_i\lb{v_i}A_z^2\rb
+\lambda\partial_i\lb\partial_i A_z^2\rb
-2\lambda\lb(\nabla{A_z})^2\rb =\\
-2\lambda\lb(\nabla{A_z})^2\rb, 
\end{array}
\label{z571}
\ee
where the last equality again follows from ignoring
 the divergence terms for suitable boundary
conditions. This  shows that $A_z$, like $B_z$ also ultimately decays.
The ultimate decay of both $B_z$ and $A_z$ implies
that the total magnetic energy also eventually decays.
The asymptotic steady-state is one of negligible total 
magnetic energy, $\lb B^2\rb \simeq 0$.   
This is the anti-dynamo theorem for 2-D (see also Moffatt 1978).  

It is important in this derivation that we took $\bfB=\curl \bfA$
because the first term on the right hand side of (\ref{z571})
would not have emerged as a straightforward total 
divergence if we were considering
a periodic box that included a 
strictly uniform $k=0$ 
periodic field that has no well defined vector potential.
(This latter case, was first studied numerically by Moss (1970)).

Now let us take advantage of the positive nature
of the last term  of  Eq.(\ref{z571}) 
to rewrite that equation
\beq
\partial_t\lb{ A_z}^2\rb=-2\lambda\lb{A_z}^2\rb/\delta^2(t),
\label{z572}
\ee
where $\delta(t)$ represents the dominant time dependent 
characteristic variation scale of $A_z$. 
Allowing for the  variation of this scale is important because 
although the magnetic energy ultimately decays in 2-D, 
random walk field line stretching can temporarily 
amplify the field. How much amplification takes place before
the field decays?

To answer this, consider  that in 2-D, the field
initially grows exponentially via the first term on the right hand side
of $(\ref{z5701})$, much like in 3-D (Kazanstev 1968; Parker 1979, Kulsrud \& Anderson 1992). Early in  the kinematic regime, 
\beq
B(t)\simeq B_0 \exp(t/\tau), 
\label{z5703a}
\ee
where $B_0$ is the root mean square of the initial magnetic field  
in the $x$--$y$ plane,
and $\tau$, for sufficiently steep kinetic energy spectra,
is the correlation time of the forcing scale. 
It should be noted that 2-D forced turbulence exhibits a steeper kinetic
energy spectrum 
than  3-D turbulence because of the tendency for enstrophy to inverse
cascade (e.g. Davidson 2004). 
Approximate inertial range energy spectra $E(k) \propto k^{-3}$ 
proposed by Batchelor (1969) Kraichnan (1970) seem to be consistent with 2-D
 simulations (Gotoh 1998; Lindborg \& Alvelius 2000)
though perhaps for slightly different reasons than originally thought
(Davidson 2004).
In 3-D, the kinetic energy spectrum $E(k) \propto k^{-5/3}$ is such that the 
the $\tau$ appearing in (\ref{9}) would
instead be closer to that of the smallest eddy scale than the forcing scale 
(Kulsrud \& Anderson 1992)).

As the field lines stretch from the turbulent motions in a confined
area, an increase in field energy means that 
the field must build up on smaller and smaller scales, so $\delta(t)$ 
decreases.
For incompressible flow, 
the field strength increases
linearly with the length of the field line. The length
of the field line varies inversely  with the scale of the
field variation. We can therefore write 
\beq
\delta(t)=L(B_0/B(t))^q =L Exp[-qt/\tau],
\label{z573}
\ee
where $q$ is an
index to allow deviations from a linear relation, 
$L$ the scale of the initial field 
$B_0$, and we have used (\ref{z5703a}). 
Combining (\ref{z572}) and (\ref{z573}) gives
\beq
\partial_t ln \lb A_z^2 \rb = -2\lambda Exp[2qt/\tau]/L^2
\label{z574}
\ee
so
\beq
\begin{array}{r}
ln ({\lb A_z^2 \rb/\lb A_z^2 \rb_0})= -(\tau\lambda/q L^2)(Exp[2qt/\tau]-1)
\\
\simeq-
\left({1\over qR_M}\right)\left(Exp[2qt/\tau]\right)=
-
\left({1\over qR_M}\right)\left({B(t)\over B_0}\right)^2,
\label{z575}
\end{array}
\ee
where we have defined the magnetic Reynolds number  
$R_{M,L}\equiv {L^2\over \lambda \tau}$, 
and where the similarity follows for  $2qt>\tau$.
Thus
\beq
\lb A_z^2 \rb=
\lb A_z^2 \rb_0 Exp[-(1/qR_{M,L})(B(t)/B_0)^{2q}].
\label{z576}
\ee
The field grows by stretching until the argument
of (\ref{z576}) becomes $\gsim 1$.
This occurs when 
\beq
B(t)\sim B_0(qR_{M,L})^{1/2q}.
\label{z577}
\ee
For the case in which $B_0={\overline B}$,
we can also write $B(t)\sim \lb b^2 \rb^{1/2}$ so   
(\ref{z577}) then gives 
\beq
\lb b^2\rb \sim {\overline B}^2(qR_{M,L})^{1/q},
\label{z578}
\ee
which for $q=1$ is the 2-D Zeldovich relation.

Using (\ref{z573}), Eq.
 (\ref{z577}) 
also implies
\beq
\delta_{m}\simeq {L\over (qR_{M,L})^{1/2}}.
\label{z580}
\ee
This is the minimum scale at which the 
field energy could peak 
in the kinematic regime 
and at which point the vector potential and the total magnetic energy would
rapidly decay. If the peak is away from this minimum
scale, then the decay rate of $A_z$ is reduced:
Using (\ref{z580}) we can rewrite (\ref{z576}) as
\beq
\lb A_z^2 \rb=
\lb A_z^2 \rb_0 Exp[-(\delta_{m}/\delta(t))^2].
\label{z576b}
\ee
At the time  Eq. (\ref{z578}) is satisfied 
the vector potential rapidly decays from (\ref{z576}) or (\ref{z576b}), 
and so does the total field energy.
The Zeldovich relation (\ref{z578}) therefore, simply 
provides an estimate of the maximum value that the fluctuating field energy could
obtain if it were to grow kinematically to this maximum
(see also \cite{pw78}).
There is no steady state for which (\ref{z578}) applies.
Moreover, (\ref{z578}) is a kinematic result
relevant for $\lb v^2\rb > R_{M,L} B_0^2$.
If instead $\lb v^2\rb < R_{M,L} B_0^2$, 
the dynamical amplification of $\lb b^2\rb$ is ultimately limited
by near equipartition with the kinetic energy density 
well before the value implied by (\ref{z578}) is reached.
This latter statement also applies even when $\bbB$ represents
a $k=0$ component; while the above proof of the total 
magnetic energy decay would not be valid in this case, 
equipartition with the kinetic energy would determine the limiting
magnetic energy.


Note also that the mean field enters (\ref{z578}) only
if the initial field energy is determined by
 $\bbB$.  The same calculation would go through even if there were 
no initial mean field, in which case the initial field 
$B_0$ would not be related to $\bB$.

\section{Understanding the 3-D Zeldovich Relation}

In 3-D, unlike 2-D, 
sustained dynamo action is allowed when the system
is forced with a velocity flow, 
whether or not there is a $k=0$ mean field.
As discussed in the introduction, simulations of incompressible MHD turbulence
show that the  magnetic field typically grows to approach and saturate 
near equipartition with  the turbulent kinetic energy in
the steady state, independent of any weak initial large scale or $k=0$ 
mean field.
In this section we show why the generalized Zeldovich relation for 
3-D does not  account for this steady-state correctly.
The distinction between large scale $0 < k < k_f$ vs. mean $k=0$
is not as essential for 3-D as it was in 2-D in what follows.

\subsection{Deriving the 3-D Zeldovich relation}

\subsubsection{The steady-state Z83 approach}

We  first follow   
Z83 and employ a spectral formalism where the 
 kinetic energy spectrum $E(k)$ is defined by
\beq
\int E(k)dk=\lb v^2\rb/2
\label{3}
\ee
and satisfies 
\beq
2E(k)=v(k)^2/k 
={v^2(k_l)\over k_l}  \left({k\over k_l}\right)^{-p},
\label{kom}
\ee
where $p$ is the spectral index, $k_l$
is the wavenumber of the turbulent forcing scale,
and $v(k)$ is the turbulent speed at wave number $k$.
Assuming  a constant  energy transfer rate
gives
\beq
v(k)^2/\tau(k)={constant},
\label{4}
\ee
so the energy transfer time $\tau(k)$ satisfies
\beq
\tau(k)
=\tau(k_l) \left({k\over k_l}\right)^{1-p}.
\label{5}
\ee
The magnetic energy spectrum $M(k)$, is defined by 
\beq
\int_{k_l}^{k_\lambda} M(k)dk= E_M
\label{6}
\ee
where $k_\lambda$ 
is the resistive wave number and 
 $E_M$ is the magnetic energy density.

Using $\bfB=\bbB+\bfb$, where $\bbB$ is a fixed large scale 
field,  and subtracting the mean of the magnetic induction equation (\ref{z5701})
from Eq. (\ref{z5701})
gives the following 
equation for the fluctuating component of the magnetic field: 
\beq
\partial_t {\bfb}=\curl(\bfv\ts \bbB)
+\curl(\bfv\ts \bfb)
-\curl\lb\bfv\ts \bfb\rb + \lambda\nabla^2 \bfb.
\label{1f}
\ee
We dot this equation with $\bfb$ and average to find
\beq
{1\over 2}\partial_t \lb b^2 \rb=
\lb\bfb\cdot\curl(\bfv\ts \bbB)\rb+
\lb\bfb\cdot\curl(\bfv\ts \bfb)\rb
+ \lambda \lb\bfb\cdot\nabla^2 \bfb\rb.
\label{2}
\ee
Now we assume that the left hand side of (\ref{2}) is zero, thus assuming that 
a steady-state is maintained by the terms on the right side.
Let us ignore the last term and non-local interactions between $\bfv$ and 
$\bfb$,  
and follow the Z83 estimates of approximate
magnitudes for the 1st and second terms on the
right. To order of magnitude, the narrow band integrated 
Fourier spectrum  for the first  term on the right is given by
\beq
\lb\bfb\cdot\curl(\bfv\ts \bbB)\rb_{(k)}
\simeq b(k) k v(k){\bB}.
\label{7}
\ee
Z83 replaces the narrow band integrated spectrum of the third term 
of (\ref{2}) with a turbulent diffusivity, that is 
\beq
\lb\bfb\cdot\curl(\bfv\ts \bfb)\rb_{(k)}
=-\nu_T(k) k^2 b^2(k), 
\label{8}
\ee
where 
\beq
\nu_T(k)\simeq {1\over \tau(k)k^{2}}.
\label{vis}
\ee

Assuming  a steady state,  setting
Eq. (\ref{7}) equal to minus Eq. (\ref{8}) gives
\beq
v(k){\bB}=\nu_T(k) k b(k). 
\label{12}
\ee
Squaring this gives
\beq
b^2(k)=2kM(k)={v^2(k){\overline B}^2\over \nu^2_T(k) k^2}=
{2E(k){\overline B}^2\over \nu^2_T(k) k}.
\label{13}
\ee
Then using (\ref{kom}), (\ref{5}) and (\ref{vis}) we have
\beq
M(k)
={{\overline B}^2\over k_l}\left({k\over k_l}\right)^{4-3p}.
\label{14}
\ee
Integrating the steady state magnetic energy spectrum 
from $k_l$ to the viscous wave number $k_\nu$  then gives  
\beq
E_{M,\nu}=
\int_{k_l}^{k_\nu} 
M(k)dk = {\overline B}^2\int_1^{k_\nu/k_l}\sigma^{4-3p}d\sigma.
\label{15}
\ee
The microphysical viscosity $\nu$ must equal
\beq
\nu=\nu_T(k_\nu)={1 \over \tau(k_\nu)k_\nu^2}
={v(k_1)\over k_l}\left({k_\nu\over k_l}\right)^{p-3},
\label{9}
\ee
where we have used (\ref{kom}) and (\ref{5}).
This implies 
\beq
R_{M,l}= Pr_M{v(k_l)\over k_l\nu}=Pr_M\left({k_l\over k_\nu}\right)^{p-3},
\label{10}
\ee
or
\beq
k_\nu=k_1 \left({R_{M,l}\over Pr_M}\right)^{1/(3-p)},
\label{11}
\ee
where $R_{M,l}\equiv{v(k_l)\over k_l\lambda}$ is the  magnetic Reynolds number associated
with scale $l$.

Let us now focus on the case $Pr_M=1$ to make contact 
with previous work.  In this case, 
$E_M=E_{M,\nu}=\int_{k_l}^{k_\nu}
M(k)dk$.
Then using (\ref{11}) in (\ref{15}) gives, for $p=5/3$
(Kolmogorov spectrum), 
\beq
\lb b^2\rb=2\int_{k_l}^{k_\nu} M(k)dk = 2{\overline B}^2 ln (k_d/k_1)={3\over 2}{\overline B}^2 ln R_{M,l}
\label{16}
\ee
For $p\ne 5/3$ we instead find
\beq
\lb b^2\rb = {{2\overline B}^2 \over 5-3p}\left[\left({k_d\over k_1}\right)^{5-3p}-1\right]\simeq
{{2\overline B}^2\over 5-3p}R_{M,l}^{5-3p\over 3-p},
\label{17}
\ee
for $k_d \gg k_l$. Eqs. (\ref{16}) and (\ref{17})
are the generalized Zeldovich relations of Z83.

Note  that for  $p=1$ 
the 
result Eq. (\ref{17}) would then imply 
\beq
\lb b^2\rb \simeq R_{M,l}{\overline B}^2, 
\label{17aa}
\ee
as noted by Z83, a formula   consistent with the 2-D result (\ref{z578}) for 
$L\sim l$.  However, as we discussed in section 2, 
there is no true steady state in 2-D and the derivation
of (\ref{17}) relied on the steady state balance between
the first two terms on the right of (\ref{2}).  Moreover, 
as also discussed below Eq. (9), the 2-D turbulent kinetic 
energy spectrum is more consistent with $p=3$, a value
steeper than the Kolmogorov value $p=5/3$, not more shallow.

Before discussing the problems with the 3-D derivation of Z83 and
the implications, we summarize 
a quite different derivation presented in KR that 
also gives a different result.

\subsubsection{The time dependent  KR approach}

KR (p.105; see also Low 1972) also 
start  with (\ref{1f}) with a fixed $\bbB$, 
but instead of replacing the
nonlinear terms by a diffusion term as in (\ref{8}), 
they employ the quasilinear approximation 
(= second order correlation or first order smoothing approximation (FOSA)) to drop the nonlinear
terms altogether.  Furthermore, they keep the $\lambda$ dissipation term and
the time evolution term on the left hand side of (\ref{8})---both of which were
dropped in the approach of the previous section.
They then Fourier transform (\ref{8}) in both
space and time to obtain
\beq
{\tilde b}_i
={\ep_{ijk}\ep_{klm}\bB_m i k_j {\tilde v}_l
\over -i\omega +\lambda k^2},
\label{rig}
\ee
where the tilde indicates Fourier transform.
Eq. (\ref{rig}) facilitates a  rigorous relation
between magnetic and velocity two-point correlators. 
After some algebra, for isotropic, incompressible turbulence this elegantly 
leads to (KR)
\beq
\lb b^2\rb = f(R_M) \bB^2
\label{39}
\ee
where 
\beq
f(R_M)
={1\over 3}\int\int {k^2 Q_{ll}({\bf k},\omega)\over \lambda^2k^4+\omega^2} 
d{\bf k}d\omega,
\ee
and
$Q_{ij}({\bf k},\omega)\delta ({\bf k}+{\bf k'},\omega +\omega 
')\equiv\lb {\tilde v}_i({\bf k},\omega){\tilde v}_j({\bf k}',{\omega}')\rb$.
Defining $s\equiv\omega/\lambda$ and taking the $\lambda \rightarrow 0$ limit, this gives 
\beq
f(R_M)\simeq {\pi \over 3\lambda}\int {Q_{ll}({\bf k},0) }d{\bf k}ds
\sim {\nu_T\over \lambda} = R_M.
\label{40}
\ee
For $\lambda\rightarrow \infty$, KR instead  obtain
$f=R_M^2$. 

The KR approach produces a different (and spectra independent) 
result for $f(R_M)$ from  the Z83 approach of the previous subsection
because KR keeps the time
derivative term in (\ref{1f}) and the resistive term, and drops
the nonlinear terms. In contrast, Z83 drops the time derivative term and
the resistive term and replaces the triple correlations by a turbulent
diffusivity. If KR were to include the nonlinear terms by a
turbulent diffusivity, the effect would be to  replace 
 $\lambda$ in this subsection  by a turbulent diffusion
coefficient.  The strong $R_M$ dependence in (\ref{40}) would
disappear, and the result would be spectrally dependent.

\subsection{Invalidating the 3-D Zeldovich relations}

Despite their differences, both approaches  to deriving
3-D Zeldovich relations in the previous two subsections
are missing the same key 
ingredient which ultimate invalidates them for realistic steady states.
Both approaches do not include the growth of the small
scale field from the third term in (\ref{1f}). 

Fundamental to KR approach, was ignoring 
any contribution from the third term in (\ref{1f}) via
the second order correlation approximation or FOSA.
But this  excludes all the nonlinear terms that make a nonlinear turbulent
spectrum. These terms are not ignorable and 
as soon as $\bfb$ grows at all beyond the fixed $\bbB$, they
become dominant.

While the  Z83 approach of section 3.1.1 does not ignore
the third term in (\ref{1f}), it treats this term 
as a diffusion term, and incorrectly assumes that a steady-state arises 
between the first two terms on the right of (\ref{2}). 
This is the central problem with the derivation 
of section 3.1.1 and invalidates (\ref{17}) for realistic steady MHD
turbulence.
As discussed earlier, 
for steady state MHD turbulence initiated with a weak
field, 
the small scale field can build up even in the absence of
a large scale field.
The saturated magnetic energy thus has nothing to do with
a mean field, and  Eq. (\ref{17}) is ruled out for the steady state.
This discrepancy between (\ref{17}) and  
numerical simulations of incompressible  MHD turbulence
(e.g. Maron \& Blackman 2002; 
Haugen et al. 2003,2004; Maron et al. 2004) 
arises because   the third term 
of (\ref{2}) 
is a actually a source of both growth and decay (e.g. 
Kulsrud \& Anderson 1992;  Haugen \& Brandenburg 2004).
The third term of (\ref{17})  and should therefore be replaced
by the sum of a growth term and  a decay term. That is, 
\beq
\lb\bfb\cdot\curl(\bfv\ts \bfb)\rb_{(k)}
=\Gamma b^2(k) -\nu_T(k) k^2 b^2(k), 
\label{8.2}
\ee
where $\Gamma\sim v(k)k$ indicates a growth rate.
For a given $k\gg k_\lambda$, 
the terms on the right can largely balance independent
of any term with the mean field or magnetic diffusivity. 
Accordingly, 
the energies of $v$ and $b$ could equilibrate without reference to 
the second term in (\ref{2}) which contains $\bbB$.
If the only initial seed field
were a weak mean field, then for a very short time,
the second term of (\ref{2}) would be the dominant
term. For a $k=0$ mean field in a box with periodic boundaries, no change
in the mean field is possible. Thus the second term of (\ref{2})
would swiftly become subdominant for an initially weak mean
field, in contrast to that implied by
the crucial step Eq. (\ref{8}), which appears in Z83.

If instead of being a weak seed, the initial $\bB$ were imposed to be 
a fraction of order unity of equipartition
with the turbulent kinetic energy, then  the first term on the right
of (\ref{2}) can be always competitive with the remaining terms on the right
of (\ref{2}).  In this case, the small scale dynamo action is modestly
suppressed by the strong mean field (Figs. 2 and 6 of 
 Haugen \& Brandenburg 2004; also Brandeburg 1993).

There is another subtlety in the derivation of (\ref{17}).
Expanding out the term containing
$\overline {\bf B}$ in (\ref{2}) gives
\beq
\lb{\bf b}\cdot\curl(\bfv\ts \bbB)\rb=\lb b_q \partial_n v_q\rb \bB_n
-\lb b_n v_q\rb \partial_q \bB_n.
\label{2.1} 
\ee
If correlations involving arbitrary products of
small scale velocity and magnetic fluctuations 
were isotropic (that is, if the velocity and magnetic fluctuations
were jointly distributed isotropically),
 both of the  terms on the right would vanish: 
the first because the correlation is a vector, and 
the second because isotropy would make that term proportional to $\delta_{qn}$.
The value of the terms on the right hand side of (\ref{2.1})
for non-helical flows and a weak mean field is therefore
less than a simple order of magnitude estimate of
those terms would indicate.  Accordingly, 
the second term of (\ref{2}) will be  smaller than 
(\ref{7}) would indicate. Symbolically, Eq. (\ref{7}) should be replaced by
\beq
|\lb\bfb\cdot\curl(\bfv\ts \bbB)\rb_{(k)}|
\simeq |b(k) k v(k)|_{a}
{\bB},
\label{7.1}
\ee
where the subscript 
$a$ indicates that only an anisotropic part of the correlation
contributes.
This issue does not arise for the  
second term on the right of (\ref{2})
(which was taken to balance the left side of (\ref{7.1}) in (\ref{12})) which is 
given by
\beq
\lb{\bf b}\cdot\curl(\bfv\ts \bfb)\rb=\lb b_q b_n \partial_n v_q \rb 
-\lb b_n v_q \partial_q b_n\rb.
\label{2.12} 
\ee
Although the second term on the right of (\ref{2.12})
vanishes for incompressible turbulence with the joint isotropy assumption, 
as can be seen by pulling the divergence out of the average, 
the first term on the right of (\ref{2.12}) has a surviving isotropic 
contribution.  Thus,  the fact that (\ref{2.1}) 
might be even smaller than estimated in (\ref{7}), further highlights 
that the terms on the right of  (\ref{8.2}) 
could largely  maintain the steady state of (\ref{2})
  by themselves.

\section{Conclusion}

We have re-examined  the meaning and derivation of the Zeldovich 
relations (Z57, Z83, ZR) of MHD for 2-D and 3-D.
These relations 
(e.g. Eqs. (\ref{16},\ref{17},\ref{17aa} \ref{39})) have been
purported to relate the fluctuating magnetic field energy to a large
scale or mean magnetic field energy in steady state forced MHD turbulence
with an imposed mean field. 
We have identified why these relations do not apply
for realistic steady states and highlight some common misconceptions
about them.

First, we have shown that these relations focus on the response
of the total magnetic energy to non-helical turbulence driven by
kinetic energy in the presence of an initially weak large scale  field.
Under the circumstances of their standard
derivations, the large scale field serves only as a seed field, and does
not grow. The relations say nothing about how large a mean
field can grow with respect to a fluctuating field in 
helical MHD turbulence and therefore place no constraints on mean field
helical MHD dynamo theory.

Second, we point out that the presence of a mean field as a seed field
in 3-D is the only reason that the small scale field that results
from MHD turbulence would depend on it. Were the initial field
a small scale seed field with no net mean, the small scale field
would still grows, as seen in nonhelically driven MHD turbulence simulations 
(e.g. Maron \& Blackman 2002; Haugen et al. 2003,2004;
Maron et al. 2004) 
Schekochihin et al. 2004), and the saturation value could not possibly 
depend on the mean
field.  For  3-D, we have indeed 
shown that standard derivations of the Zeldovich
relation do not include (KR)
 or replace key nonlinear terms in the magnetic energy
equation  by a  turbulent diffusion term (Z83), ignoring
an equally important growth term. The 
Zeldovich relations that emerge  therefore result from an inappropriate 
balance of  non-dominant terms.  Accordingly, as seen in 3-D simulations,
the ultimate steady-state 
saturation of the small scale field is determined by equipartition
with the turbulent kinetic energy, independent of
the mean field or magnetic Reynolds number for large $R_M$ non-helical systems.

Finally, for 2-D, we have shown that when 
the magnetic field has no $k=0$ component (but can still have
a large scale $k<k_f$ component) the net magnetic energy must
decay for a forced 2-D MHD turbulence system. Then the Zeldovich relation 
(\ref{z578}) or (\ref{17aa})  cannot
represent a true steady state.
We have shown that instead, this relation represents the maximum that the
fluctuating field could obtain if the initial seed field were a
large scale seed field (and if $R_M \bB^2$ were less than the turbulent
kinetic energy---otherwise the latter determines a quasi-steady saturated value) after which the magnetic energy would decay.  In this respect, the import of 
the Zeldovich relation of Z57, was mainly 
to illustrate quantitatively, the now well known
result that 2-D MHD turbulence does not sustain a steady-state dynamo
without a fixed seed.

\acknowledgements
EGB acknowledges support from 
NSF grant AST-0406799 and NASA grant ATP04-0000-0016,
and the Isaac Newton Institute, Cambridge, 
and A. Brandenburg \& M. Proctor for comments and discussions.

\end{document}